\documentclass[trackchanges,preprint2]{aastex7}

\begin{document}

\title{The Hollyfeld Gambit in Astrophysics}

\begin{abstract}
We estimate the Hollyfeld Gambit for the Powerball lottery and its return on investment compared to present and extrapolated federal funding for astrophysical grants. Using a Monte Carlo estimation of rate of return for the Powerball, we conclude a Hollyfeld Gambit is a better bet than a federal grant by the end of the decade if current trends hold. 
\end{abstract}



\section{Introduction}
\label{s:intro}

Astrophysics  is facing an increase in the odds for funding in the United States. 

Private funding has been held up as a possible alternative to federal funding grants. Because this is now constrained by the limits of philanthropy as well, alternatives need to be sought. 

In this constrained environment, we explore the viability of Lazlo Hollyfield Gambit as an alternative to federal grants and personal philanthropy.

\section{Lazlo Hollyfield Gambit}
\label{s:data}

Lazlo Hollyfeld Gambit is a Game Theory strategy to maximize entries in order to increase odds of return (Hollyfeld, 1985). This Gambit is based on an original conjecture of Klein, Novikoff and Barry Megdal (1974). The gambit is to participate in a high-return lottery, discounting personal hours invested. 

The modified Hollyfeld Gambit is to work an hourly wage job to convert into Powerball tickets. The US federal minimum wage is 7.25\$/hr. A Powerball ticket is \$2 for a standard play and \$3 including the ``power-up''. In effect, for each hour of work, one can play the Powerball lottery $3\times$. For a postdoc salary ROI, we assume a 45\$/hr salary. 

\begin{figure}
    \centering
    \includegraphics[width=\linewidth]{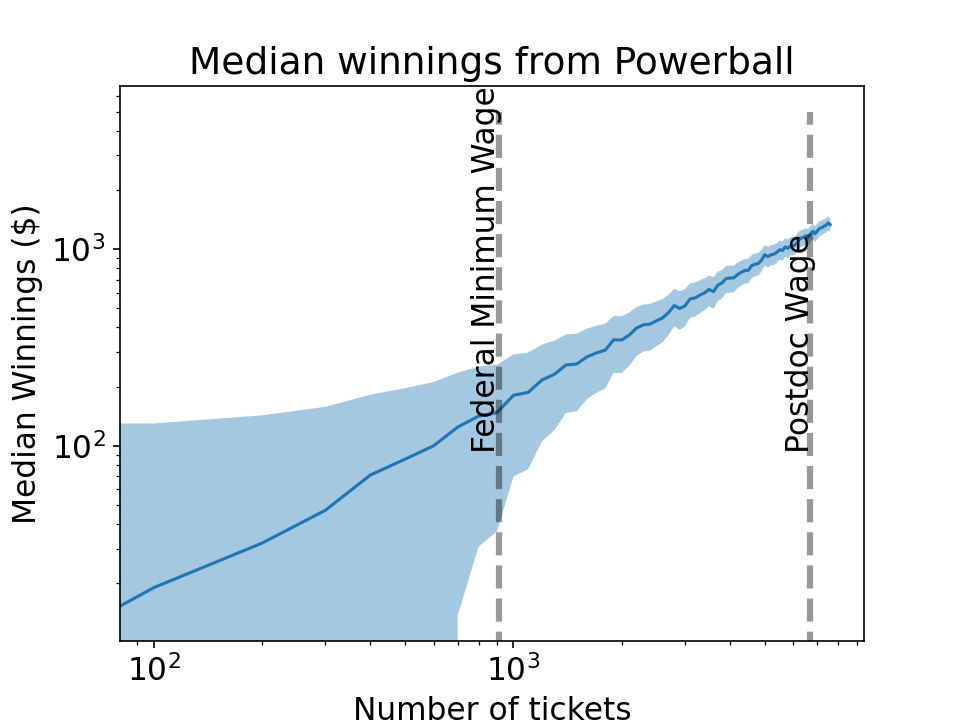}
    \caption{The median return of the Powerball as a function of number of tickets bought. The number of tickets that can be bought with 38 days of labor for a postdoc and a federal minimum wage are marked with dashed lines. }
    \label{f:powerball}
\end{figure}

In order to determine the median return on Powerball tickets played, we calculate the combined return:
\begin{equation}
    ROI = \Sigma P_i \times R_i
\end{equation}
where $P_i$ is the probability ($P_i = 1/odds$) to win a particular prize and $R_i$ is the return of that prize, i.e. the won dollar amount (\href{https://www.powerball.com/powerball-prize-chart?utm_source=chatgpt.com}{Powerball Odds}). Included is the odds for winning no prize, for which the return is negative, the prize paid for the ticket. 

\subsection{Monte Carlo Modeling}

We have modeled the Hollyfeld Gambit for each number of tickets paid using a 100 iteration monte carlo to estimate the mean and standard deviation in each. This is the blue line and shaded area in Figure \ref{f:powerball}: monte carlo median return for the number of tickets played.

\section{Federal funding}
\label{s:fedfunding}

Federal funding for the National Science Foundation's directorate of Mathematics and Physical Sciences stands in for our model for the ROI of federal grant funding (\href{https://tableau.external.nsf.gov/views/NSFbyNumbers/Trends}{NSF by the Numbers}).

\begin{figure}
    \centering
    \includegraphics[width=\linewidth]{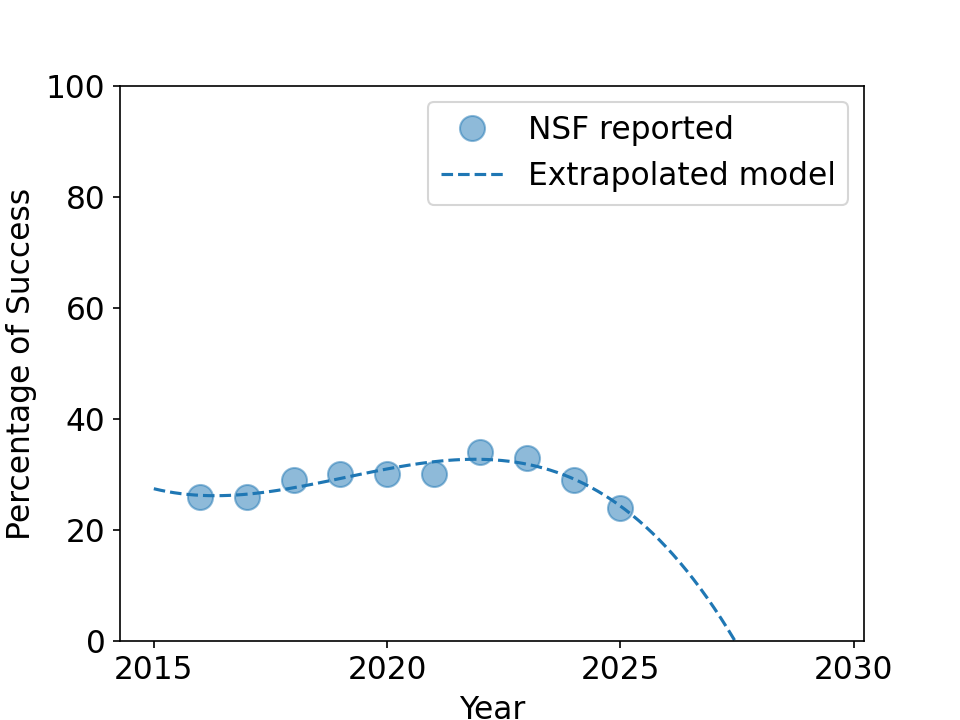}
    \caption{The percentage of grants awarded by the NSF mathematics and physical science for each year and the third degree polynomial trend.}
    \label{f:NSF:odds}
\end{figure}

Figure \ref{f:NSF:odds} shows the success rate of NSF proposals in this directorate over the period 2015-2026 and extrapolated success rate based on a third degree polynomial. This is our extrapolated model for the near future odds.

% NSF ROI
To calculate the NSF ROI, we use a similar approach as the Powerball ROI: ROI = grant \$ $\times P$, where $P$ is the odds of the grant.

We assume it takes 38 days to write, budget, shepherd past university bureaucracy and submit for each proposal (\href{https://medium.com/deip/grants-applications-writing-time-consuming-but-not-always-rewarding-d7078da088f3}{Deip 2018}). This is how we calculate the minimum wage and postdoc wage inlay that goes into  Figure \ref{f:powerball} to calculate the median return for the Powerball return. This return is then plotted in \ref{f:NSF:odds}. Figure \ref{f:NSF:odds} shows the rate of return for federal grant writing. 

The projected ROI of federal grants intersects in 2029 with the Powerball returns for the same effort --38 days--  for both postdoc and minimum wage.

\begin{figure}
    \centering
    \includegraphics[width=\linewidth]{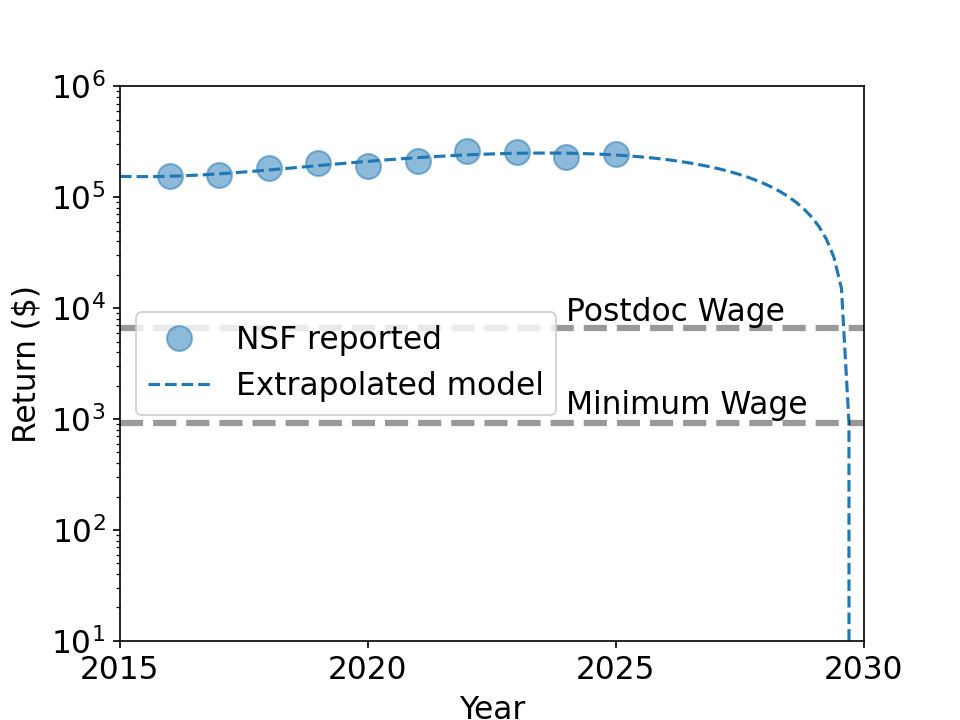}
    \caption{The ROI for NSF data and projection and the Powerball returns from Figure \ref{f:powerball}. }
    \label{f:NSF:roi}
\end{figure}

\section{Discussion}
\label{s:discussion}

Prognosticating on the best use of researcher's effort in the coming years, it appears that the trend in Figure \ref{f:NSF:roi} favors the Hollyfeld Gambit before the end of the decade. Our basic assumptions are that a federal proposal represents approximately 38 days of work and that the ROI trend line in Figure \ref{f:NSF:roi} continues to hold. In the near future, some of these assumptions may change. The trend may be reversed, the odds of the Powerball may change, the Federal minimum wage may change, although that has proven stable for the past 17 years, 

If the trend in Figure \ref{f:NSF:roi} does hold, it seems likely that the Hollyfeld Gambit will pay off to early adopters, those who switch to it before the break even point.

\section{Conclusions}
\label{s:conclusions}

We evaluated the Hollyfeld Gambit (1984) for the Powerball lottery in the context of current trends of Federal ROI on grant writing efforts. We conclude that it will be cost-effective to attempt the Hollyfeld Gambit rather than a federal grant opportunity by the end of the decade, regardless of the pay grade of the PI.

%\bibliography{Bibliography}{}
%\bibliographystyle{aasjournalv7}
\section*{References}

Klein, Novikoff and Barry Megdal, MIT prank, 1974

Hollyfeld, Tri-Star Pictures, 1985

% 

% \url{https://en.wikipedia.org/wiki/On_Bullshit}



\end{document}